\documentclass[notitlepage,aps,nofootinbib,tightenlines,11pt,letterpaper,prl,twocolumn,amsmath,amssymb,amsfonts,long,superscriptaddress,preprintnumbers]{revtex4-1}
\pdfoutput=1

\usepackage{fixmath}
\usepackage{appendix}
\usepackage{enumitem}
\usepackage[dvipsnames]{xcolor}
\usepackage{amsmath}
\usepackage{mathtools}
\usepackage[utf8]{inputenc}
\usepackage[english]{babel}
\usepackage{graphicx}
\usepackage{epsfig}
\usepackage{slashed}
\usepackage{amssymb}
\usepackage{color}
\usepackage{tabulary}
\usepackage{multirow}
\usepackage[normalem]{ulem}
\definecolor{MyDarkBlue}{rgb}{0.1, 0.1, 0.8} 
\definecolor{SBlue}{rgb}{0.2, 0.4, 0.7} 
\definecolor{MyLightBlue}{rgb}{0.22,0.51,0.9}
\definecolor{MyGreen}{rgb}{0.0, 0.5, 0.0}
\definecolor{BrickRed}{rgb}{0.8, 0.25, 0.33}
\RequirePackage{hyperref}
\hypersetup{colorlinks, citecolor=SBlue,linkcolor=MyDarkBlue, urlcolor=PineGreen}
\makeatletter
\usepackage{amssymb}
\usepackage{pifont}

\newcommand{\prlsection}[2]{\noindent{\it\textbf{#1}{#2}}---}
\makeatletter
\newcommand*{\balancecolsandclearpage}{%
	\close@column@grid
	\cleardoublepage
	\twocolumngrid
}
\makeatother

\DeclareUnicodeCharacter{2212}{-}
\DeclareUnicodeCharacter{2217}{-}
\DeclareUnicodeCharacter{2113}{-}

\begin{document}

\title{\Large 
Probing the $\mathbold{R_{K^{(*)}}}$ Anomaly at a Muon Collider}

\author{Guo-yuan Huang}
\email[E-mail: ]{guoyuan.huang@mpi-hd.mpg.de}
\affiliation{Max-Planck-Institut f{\"u}r Kernphysik, Saupfercheckweg 1, 69117 Heidelberg, Germany}

\author{Sudip Jana}
\email[E-mail: ]{sudip.jana@mpi-hd.mpg.de}
\affiliation{Max-Planck-Institut f{\"u}r Kernphysik, Saupfercheckweg 1, 69117 Heidelberg, Germany}

\author{Farinaldo S.\ Queiroz}
\email[E-mail: ]{farinaldo.queiroz@iip.ufrn.br}
\affiliation{International Institute of Physics, Universidade Federal do Rio Grande do Norte,
Campus Universitario, Lagoa Nova, Natal-RN 59078-970, Brazil}
\affiliation{Departamento de Fisica, Universidade Federal do Rio Grande do Norte, 59078-970, Natal, RN, Brasil
Millennium Institute for SubAtomic Physics at the High-energy frontIeR, SAPHIR, Chile}

\author{Werner Rodejohann}
\email[E-mail: ]{werner.rodejohann@mpi-hd.mpg.de}
\affiliation{Max-Planck-Institut f{\"u}r Kernphysik, Saupfercheckweg 1, 69117 Heidelberg, Germany}

\begin{abstract}
\noindent 
The LHCb measurements of the $\mu / e$ ratio in $B \to K \ell \ell$
decays $(R_{K^{}})$ indicate a deficit with respect to the Standard Model prediction, supporting earlier hints of lepton universality violation observed in the $R_{K^{(*)}}$ ratio. Possible explanations of these  $B$-physics anomalies include heavy $Z'$ bosons or scalar and vector leptoquarks mediating $b \to s \mu^+  \mu^-  $. 
We note that a muon collider can directly measure this process via $\mu^+  \mu^- \to b \bar s$ and can shed light on the lepton non-universality scenario.  Investigating currently discussed center-of-mass energies $\sqrt{s} = 3$, 6 and 10 TeV, we show that the parameter space of $Z'$ and leptoquark solutions to the $R_{K^{(*)}}$ anomalies can be mostly covered. Effective operators explaining the anomalies can be probed with the muon collider setup $\sqrt{s} = 6~{\rm TeV}$ and integrated luminosity $L = 4~{\rm ab^{-1}}$.

\end{abstract}

\maketitle

\prlsection{Introduction}{.}%
Rare decays of mesons are sensitive to effects of heavy particles. Precision studies of many such decays have confirmed the CKM matrix as the source of flavor transitions in the Standard Model (SM) \cite{Zyla:2020zbs}. Nevertheless, long-standing hints for physics beyond the CKM paradigm exist. In particular, decay rates of charged and neutral $B$ mesons into kaons plus first and second generation charged leptons are notoriously away from precisely known SM calculations by $3.1\sigma$~\cite{Aaij:2021vac, Aaij:2017vbb,Aaij:2019wad}. A straightforward solution to these so-called $R_{K^{(*)}}$ puzzles is that there is new physics in the transition $b \to s \mu^+ \mu^-$, which can be rewritten as $\mu^+ \mu^- \to b \bar s $. For energy scales of $B$ decays this physics can be described by effective operators, which may stem from heavy particles mediating the transition. Essentially, there are only two possibilities at tree level. New $Z'$ bosons that couple to $b\bar s$ and $\mu^+ \mu^-$ 
or hypothetical leptoquarks that couple to $\mu^- b$ and $\mu^+ \bar s$.

This paper is about realizing the process $\mu^+  \mu^- \to b \bar s (\bar b s)$ at high energy muon colliders. Those are currently under active
discussion~\cite{MCmeeting,MCgoal,Cheung:2021iev,Liu:2021jyc,Han:2021udl,Chiesa:2020awd,Costantini:2020stv,Han:2020pif,Han:2020uak,Bandyopadhyay:2020otm,Gu:2020ldn,Capdevilla:2021fmj,Capdevilla:2020qel,Buttazzo:2020eyl,Yin:2020afe,Huang:2021nkl,Capdevilla:2021rwo} as a possible future collider. While being interesting for Higgs physics, they would in particular be a powerful probe for anything new that likes muons. In particular, the option to test physics solutions for the anomalous magnetic moment of the muon, another long-standing problem 
involving muons \cite{Bennett:2006fi,Roberts:2010cj,Aoyama:2020ynm}, has been investigated. It has been shown that any new physics that may be responsible for explaining the $(g-2)_\mu$ results can be tested at future muon colliders \cite{Capdevilla:2020qel,Buttazzo:2020eyl,Yin:2020afe,Huang:2021nkl,Capdevilla:2021rwo}. 
Here we discuss the $B$ physics anomalies in the $R_K$ and $R_{K^*}$  ratios in terms of a $Z'$ and scalar as well as vector leptoquarks. The former mediates the process $\mu^+  \mu^- \to b \bar s$ in an $s$-channel diagram, the latter in a $t$-channel diagram. 
Using the currently discussed setups of 3, 6 and 10 TeV center-of-mass energies \cite{MCmeeting,MCgoal}, we show that both scenarios can be mostly covered. Our analysis takes di-jet background from SM processes into account, and is independent of whether flavor tagging is included or not.  
Before turning to the 
analysis at the muon collider, we will shortly summarize the current situation of the anomalies and their main solutions. 

\vspace{0.05 in} 
\prlsection{Theoretical interpretations of the $\mathbold{R_{K^{(*)}}}$ anomaly}{.}%
The ratios $R_K$ and $R_{K^*}$, relevant for testing the universality of the gauge-interactions in the lepton-sector, are defined as 
\begin{align} \label{Eqn:RK}
R_{K} \ & = \ \frac{{\rm BR}(B^+ \to K^{+} \mu^+ \mu^-)}{{\rm BR}(B^+ \to K^{+} e^+ e^-)} \;, \\ 
R_{K^{\star}} \ & = \ \frac{{\rm BR}(B^0 \to  K^{\star0} \mu^+ \mu^-)}{{\rm BR}( B^0 \to  K^{\star0} e^+ e^-)} \;.
\end{align}
Due to highly suppressed hadronic uncertainties, such ratios are supposed to be theoretically clean and could thus be  a  ‘clean’-signal of BSM-physics. 
Very recently, the LHCb collaboration reported the results  of $R_K$-measurement (in the region  $q^2\in [1.1,6]~{\rm GeV}^2$) 
as~\cite{Aaij:2021vac}
\begin{eqnarray} \label{Eqn:RK-Exp-new}
R_K^{\rm LHCb} \ = \ 0.846^{+0.042+0.013}_{-0.039-0.012}\,  ,
\end{eqnarray}
which indicates a $3.1\sigma$ discrepancy from its SM 
prediction~\cite{Bobeth:2007dw, Bordone:2016gaq} 
\begin{eqnarray} \label{Eqn:RK-SM}
R_K^{\rm SM} \ = \ 1.0003\pm 0.0001\,.
\end{eqnarray}
Similarly, the LHCb Collaboration has also reported the results of $R_{K^{*}}$-measurement in two low-$q^2$ bins~\cite{Aaij:2017vbb} ($q^2\in [0.045,1.1]~{\rm GeV}^2$ and $q^2\in [1.1,6]~{\rm GeV}^2$):
\begin{eqnarray}
R_{K^*}^{\rm LHCb}& \ = \ & \begin{cases}0.660^{+0.110}_{-0.070}\pm 0.024 \, , \\ 
0.685^{+0.113}_{-0.069}\pm 0.047  \, ,
\end{cases}
\end{eqnarray}
which shows $2.2\sigma$ and $2.4\sigma$ deviations, respectively from their corresponding SM-predictions in each $q^2$ bin~\cite{Capdevila:2017bsm, Alok:2017sui}:
\begin{eqnarray}
R_{K^\star}^{\rm SM} \ = \  \begin{cases} 0.92\pm 0.02 \, , \\  
1.00\pm 0.01\,.
\end{cases}
\end{eqnarray}
Furthermore, Belle has also presented their  results  on  $R_K$~\cite{Abdesselam:2019lab} and  $R_{K^*}$~\cite{Abdesselam:2019wac}. However, there are comparatively larger uncertainties than for the LHCb measurements.  There are in fact only a few BSM possibilities which could resolve these $R_{K^{(*)}}$-anomalies. Before entering details, it is quite important to mention that an explanation of $R_{K^{(*)}}$ by modifying  the $b\to s\mu^+\mu^-$ decay anticipates a better global-fit to other observables, as compared to altering the $b\to se^+e^-$ decay~\cite{Aebischer:2019mlg}.

The effective Lagrangian responsible for semi-leptonic $b \rightarrow s \mu^{+} \mu^{-}$-transitions can be expressed as ($V$ denotes the CKM-matrix)
\begin{equation}\label{eq:Leff}
\mathcal{L}_{b \rightarrow s \mu \mu}^{\mathrm{NP}} \supset \frac{4 G_{\rm F}}{\sqrt{2}} V_{t b} V_{t s}^{*}\left(C_{9}^{\mu} O_{9}^{\mu}+C_{10}^{\mu} O_{10}^{\mu}\right)+\mathrm{h.c.} 
\end{equation}
with the relevant operators 
\begin{equation}
\begin{aligned}
O_{9}^{\mu} &=\frac{\alpha}{4 \pi}\left(\bar{s}_{\rm L} \gamma_{\mu} b_{\rm L}\right)\left(\bar{\mu} \gamma^{\mu} \mu\right) ,\\
O_{10}^{\mu} &=\frac{\alpha}{4 \pi}\left(\bar{s}_{\rm L} \gamma_{\mu} b_{\rm L}\right)\left(\bar{\mu} \gamma^{\mu} \gamma_{5} \mu\right) .
\end{aligned}
\end{equation}
Using these operators to explain the anomalies leads to best-fit values of the Wilson-coefficients  $ C_9 = -C_{10} = -0.43$, with the $1 \, \sigma$ range being  $ [-0.50,-0.36]$~\cite{Altmannshofer:2021qrr,Aebischer:2019mlg}.

\begin{figure}[htb!]
    \centering
   \hspace{-0.1in} \includegraphics[width=.48\textwidth]{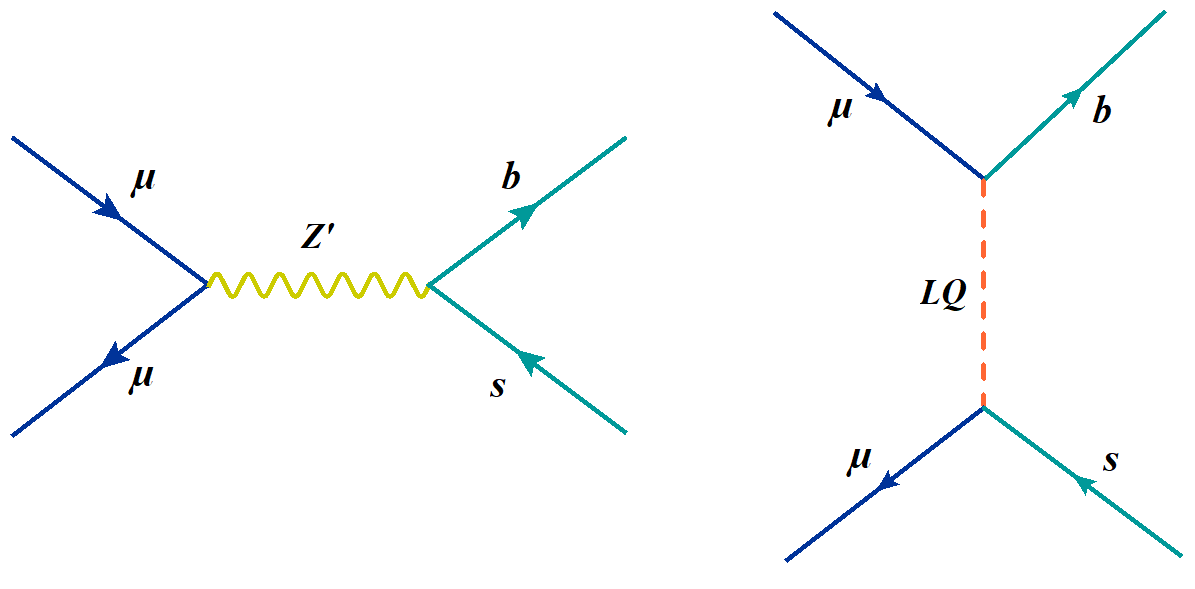} 
    \caption{Tree-level processes at a muon collider directly related to $R_{K^{(*)}}$: $Z'$ or leptoquark.
    \label{fig:feyn}}
\end{figure}

\vspace{0.05 in} 
\prlsection{Models with $\mathbold{Z'}$}{.}%
Let us now discuss an explicit new-physics realization for explaining the $B$-anomalies in neutral-currents. As a prototypical-model (a partial list of references is \cite{1310.1082, 1311.6729, 1403.1269, 1501.00993, 1503.03477, 1503.03865, 1505.03079, 1506.01705, 1508.07009, 1509.01249, Calibbi:2019lvs, 1510.07658, 1512.08500, 1601.07328, 1602.00881, 1604.03088, 1608.01349, 1608.02362, 1611.03507, 1702.08666, Kawamura:2019hxp, 1703.06019, 2009.02197}), we consider a $Z^{\prime}$ which dominantly couples to $b s$ and $\mu^{+} \mu^{-},$ via left-handed 
currents\footnote{Right-handed currents in the lepton-sector actually worsen the compatibility of  $R_{K^{(*)}}$ explanation with the  $\Delta M_{s}$ (mass-differences of neutral $B$-mesons) measurement \cite{DiLuzio:2019jyq}, since they demand a larger Wilson-coefficient.}. One can achieve this by extending the SM with an extra $U(1)$ gauge group, which brings in a new $Z^{\prime}$ boson having a non-universal lepton-coupling and a flavor-changing quark-coupling. Here, we  concentrate solely on the Lagrangian-part relevant for $b \rightarrow s \mu^{+} \mu^{-}$-transitions, namely
\begin{equation}
\mathcal{L}_{Z^{\prime}} \supset \left(\lambda_{i j}^{\rm Q} \bar{d}_{\rm L}^{i} \gamma^{\mu} d_{\rm L}^{j}+\lambda_{\alpha \beta}^{\rm L} \bar{\ell}_{\rm L}^{\alpha} \gamma^{\mu} \ell_{\rm L}^{\beta}\right) Z_{\mu}^{\prime} \;,
\end{equation}
where $\ell^{i}$  and $d^{i}$ denote the different generations of charged-lepton and  down-type quark states,  respectively.

\begin{figure}[t]
	\begin{center}
		\includegraphics[width=0.5\textwidth]{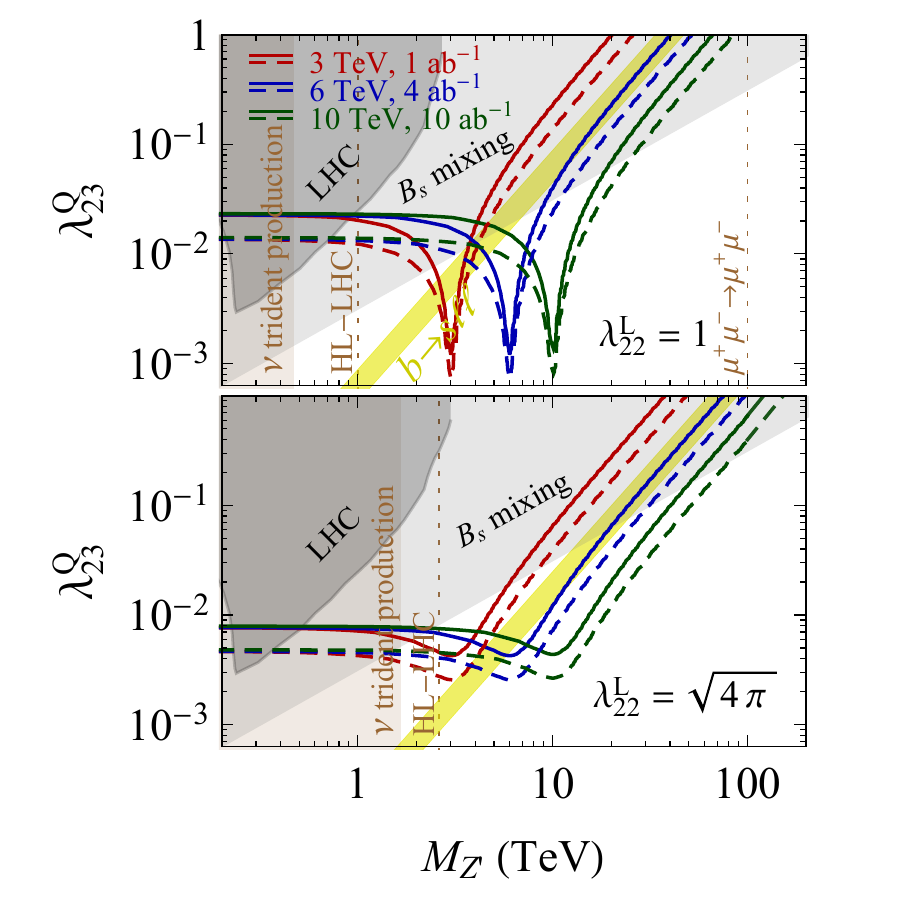} 
		\vspace{-1cm}
	\end{center}
	\caption{The sensitivity contours for the $Z'$ model with $\lambda^{\rm L}_{22} = 1$ (upper panel) and $\lambda^{\rm L}_{22} = \sqrt{4\pi}$ (lower panel) via the process $\mu^{+}\mu^{-} \to b \bar s$ at muon colliders with the following setups: 
	$\sqrt{s} = 3~{\rm TeV}$ and $L  = 1~{\rm ab^{-1}}$ (red curves), $\sqrt{s} = 6~{\rm TeV}$ and $L  = 4~{\rm ab^{-1}}$ (blue curves), as well as $\sqrt{s} = 10~{\rm TeV}$ and $L  = 10~{\rm ab^{-1}}$ (green curves). The $2\sigma$ parameter space favored by a fit of $B$ anomalies is shown as the yellow band \cite{Altmannshofer:2021qrr}. Dashed 
	(solid) curves stand for the case with (without)  flavor tagging. The $B^{}_{s}$ mixing bounds are given as gray shaded regions~\cite{DiLuzio:2019jyq}. The limits from neutrino trident production are recast as brown shaded regions~\cite{Altmannshofer:2014pba}. 
	{The regions disfavored by LHC dimuon resonance searches  are shown as black shaded regions, rescaled from Ref.~\cite{Allanach:2015gkd}. This limit is overestimated as all light quarks are assumed to couple  identically  to $Z'$.}
	The projected sensitivity of HL-LHC is given by the vertical dotted lines near $1~{\rm TeV}$~\cite{delAguila:2014soa}. {
	The $\mu^+\mu^- \to \mu^+ \mu^-$ process at the muon collider can probe all $M^{}_{Z'}$ values smaller than $100~{\rm TeV}$ with order one $\lambda^{\rm L}_{22}$~\cite{Huang:2021nkl}. These are shown as two vertical lines near $100~{\rm TeV}$.} }
	\label{fig:contours_Zp}
\end{figure}

Integrating out the $Z'$ field, one can  obtain the  effective-Lagrangian as:
\begin{equation}
\begin{aligned}
\mathcal{L}_{Z^{\prime}}^{\mathrm{eff}} &=-\frac{1}{2 M_{Z^{\prime}}^{2}}\left(\lambda_{i j}^{\rm Q} \bar{d}_{\rm L}^{i} \gamma_{\mu} d_{\rm L}^{j}+\lambda_{\alpha \beta}^{\rm L} \bar{\ell}_{\rm L}^{\alpha} \gamma_{\mu} \ell_{\rm L}^{\beta}\right)^{2} \\
& \supset-\frac{1}{2 M_{Z^{\prime}}^{2}}\left[\left(\lambda_{23}^{\rm Q}\right)^{2}\left(\bar{s}_{\rm L} \gamma_{\mu} b_{\rm L}\right)^{2}\right.\\
&\left.+2 \lambda_{23}^{\rm Q} \lambda_{22}^{\rm L}\left(\bar{s}_{\rm L} \gamma_{\mu} b_{\rm L}\right)\left(\bar{\mu}_{\rm L} \gamma^{\mu} \mu_{\rm L}\right)+\mathrm{h.c.}\right].
\end{aligned}
\end{equation}
Now one can find the relevant Wilson-coefficients at  tree-level [cf.\ left-panel of Fig.~\ref{fig:feyn}] by matching onto the effective-Lagrangians for the low-energy observables  at the scale $\left(\mu=M_{Z^{\prime}}\right)$ as
\begin{equation} \label{eq:wc_zp}
C_{9}^{\mu}=-C_{10}^{\mu}=-\frac{\pi}{\sqrt{2} G_{\rm F} M_{Z^{\prime}}^{2} \alpha}\left(\frac{\lambda_{23}^{\rm Q} \lambda_{22}^{\rm L}}{V_{t b} V_{t s}^{*}}\right).
\end{equation}
However, as shown e.g.\ in Refs.~\cite{DiLuzio:2019jyq,DiLuzio:2017fdq}, this $Z'$-explanation of  $R_{K^{(*)}}$ anomaly is under tight constraints from several theoretical and experimental limits allowing a  narrow mass-range for $Z'$ boson [cf.\ the yellow band in Fig.~\ref{fig:contours_Zp}]. There are several  dedicated $Z'$-searches  at the LHC looking at dimuon  or dijet \cite{Aad:2020otl, Beghin:2019jhu} signatures. The reliance on parton distribution functions of bottom-quarks for production in our scenario dilutes the impact of current LHC-searches. On the other hand, a very stringent bound on our $Z'$ originates from its flavor-changing coupling, which generates an additional contribution to  $B_{s}-\bar{B}_{s}$ mixing \cite{DiLuzio:2019jyq, DiLuzio:2017fdq}. Note that other constraints, such as  $\operatorname{BR}(B \rightarrow K \bar{\nu} \nu)$ \cite{Lees:2013kla} or  muon $g-2$ \cite{Bennett:2004pv,Queiroz:2014zfa},  are much weaker. In addition, there will be constraints from the  measurement of neutrino-trident production \cite{Altmannshofer:2014pba}. All these constraints are summarized in Fig.~\ref{fig:contours_Zp}.

\vspace{0.05in}
\prlsection{Models with leptoquarks}{.}%
In order to address the $R_{K^{(*)}}$-anomaly, there is another popular class of models (a partial list of references is~\cite{Hiller:2014yaa, Gripaios:2014tna, Varzielas:2015iva, Becirevic:2015asa, Alonso:2015sja, Bauer:2015knc, Fajfer:2015ycq, Barbieri:2015yvd, Becirevic:2016oho, Becirevic:2016yqi, Crivellin:2017zlb, Hiller:2017bzc, Becirevic:2017jtw, Dorsner:2017ufx, Assad:2017iib, DAmico:2017mtc, Marzocca:2018wcf, Blanke:2018sro, DiLuzio:2017vat, Calibbi:2017qbu, Bordone:2017bld, Becirevic:2017jtw, Saad:2020ucl, Babu:2020hun}) in which leptoquarks are applied.  Here we briefly review these simplified models that can accommodate the $R_{K^{(*)}}$-anomaly. There are only four scalar leptoquarks  which can interact with the SM-fermions at renormalizable level.   Interestingly, $S_3 \sim (3,3,-1/3)$  can simultaneously address $R_K$ and $R_{K^*}$ and whose constraints are not in conflict with the experimental data \cite{Queiroz:2014pra,Angelescu:2018tyl}. 
{Similarly, the vector leptoquark $U_{1} \sim (3,1,2/3)$ can also provide a good fit for the $R_{K^{(*)}}$-anomaly. Note that it requires a proper UV-completion for theoretical consistency. Here we focus mainly on the scalar case, delegating details of the vector leptoquark case to the supplemental material.}

\begin{figure}[t]
	\begin{center}
		\includegraphics[width=0.5\textwidth]{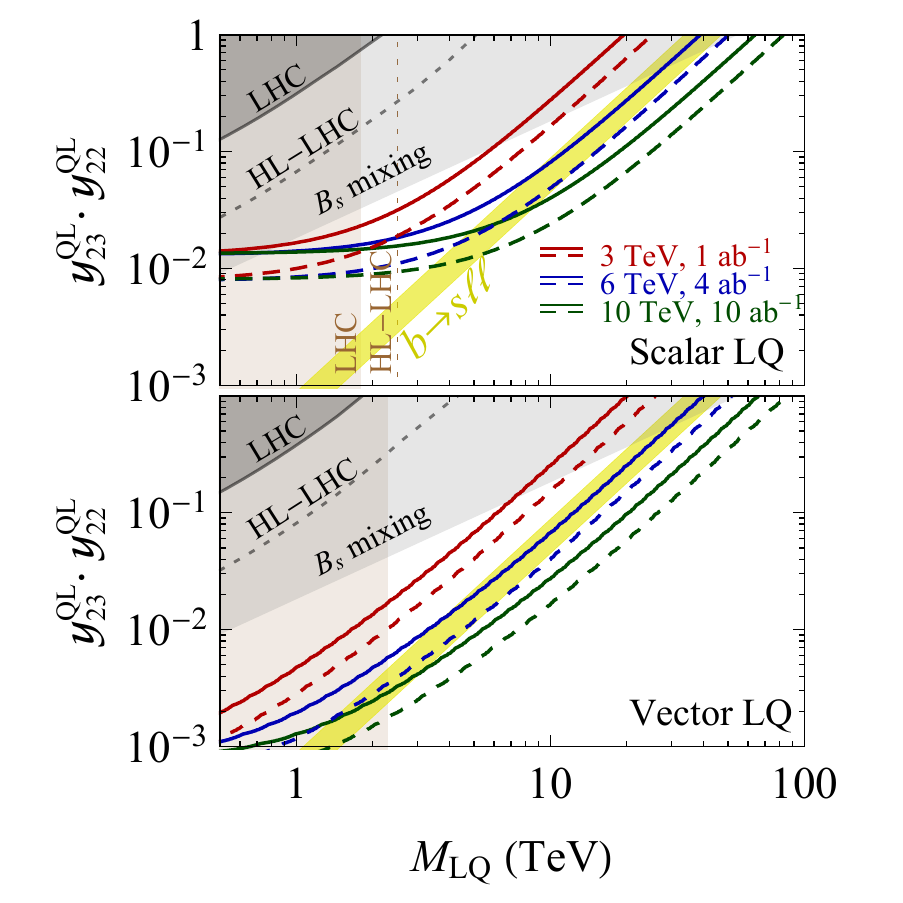} 
		\vspace{-0.5cm}
	\end{center}
	\caption{The sensitivity contours for $S^{}_{3}$ (upper panel) and $U^{}_{1}$ (lower panel) leptoquark models via the process $\mu^{+}\mu^{-} \to b \bar s$ at muon colliders with the following setups: 
	$\sqrt{s} = 3~{\rm TeV}$ and $L  = 1~{\rm ab^{-1}}$ (red curves), $\sqrt{s} = 6~{\rm TeV}$ and $L  = 4~{\rm ab^{-1}}$ (blue curves), as well as $\sqrt{s} = 10~{\rm TeV}$ and $L  = 10~{\rm ab^{-1}}$ (green curves). The $2\sigma$ parameter space favored by a fit of $B$ anomalies is shown as the yellow band \cite{Altmannshofer:2021qrr}. Dashed (or solid) curves stand for the case with (or without) the flavor tagging. The $B^{}_{s}$ mixing bound on leptoquark is given as gray shaded regions~\cite{DiLuzio:2017fdq}. The constraints by LHC searches of { leptoquark pair production (or indirect high-energy tails $q \bar{q} \to \mu^+ \mu^-$)} as well as the future projection of high-luminosity LHC~\cite{Allanach:2019zfr,Angelescu:2021lln} are given by the brown (or black) shaded region and the dotted line, respectively.
	{Note that for $U^{}_{1}$ leptoquark, the assumption $\mathcal{L} \supset \kappa g^{}_{\rm s} U_{1}^{\dagger \mu} G^{}_{\mu\nu} U_{1}^{\nu}$ with $\kappa=1$ has been made in deriving the constraints from leptoquark pair production~\cite{Angelescu:2021lln}.}}	\label{fig:contours_LQ}
\end{figure}

The relevant Lagrangian for $S_3$ can be written as:
\begin{equation}
\mathcal{L}_{S_{3}}=-M_{S_{3}}^{2}\left|S_{3}^{a}\right|^{2}+y_{i \alpha}^{\rm LQ} \overline{Q^{\rm c}}^{i}\left(\epsilon \sigma^{a}\right) L^{\alpha} S_{3}^{a}+\mathrm{h.c.},
\end{equation}
with lepton and quark-doublets $L^{\alpha}=\left(\nu_{\rm L}^{\alpha},~ \ell_{\rm L}^{\alpha}\right)^{\rm T}$ and $Q^{i}=\left(V_{j i}^{*} u_{\rm L}^{j}, ~ d_{\rm L}^{i}\right)^{\rm T}$, and Pauli-matrices   $\sigma^{a}$ ($a=1,2,3$; $\epsilon=i \sigma^{2}$). 
The leptoquark contributes to the Wilson-coefficients at  tree-level [cf.\ Fig.~\ref{fig:feyn}] and one can identify: 
\begin{equation} \label{eq:wc_lq}
C_{9}^{\mu}=-C_{10}^{\mu}=\frac{\pi}{\sqrt{2} G_{\rm F} M_{S_{3}}^{2} \alpha}\left(\frac{y_{32}^{\rm LQ} y_{22}^{\rm LQ *}}{V_{t b} V_{t s}^{*}}\right). 
\end{equation}
This explanation of the $R_{K^{(*)}}$ anomaly also faces several theoretical and experimental constraints. The same combination of Yukawa-couplings leads  to $B_{s}-\bar{B}_{s}$ mixing at one-loop level \cite{Davidson:1993qk,Dorsner:2016wpm,DiLuzio:2017fdq}. This sets an upper bound on the Yukawa-couplings as a function of the leptoquark mass as shown in Fig.~\ref{fig:contours_LQ}. Due to the loop-nature of this constraint, it is much weaker compared to the $Z'$ scenario. There are several relevant direct LHC searches. Pair-production via gluon-gluon fusion processes dominates and the subsequent decay into $\mu j$ can be looked for. A stringent limit from a dedicated LHC search using $\mu \mu j j$ signals exists \cite{Aad:2020iuy}. Recently,  Ref.\ \cite{Allanach:2019zfr} has worked out in detail the prospect of  probing the $S_3$ leptoquark at  current and future runs of the LHC. Based on that  analysis  masses up to 1.8 TeV are excluded at $95\%$ confidence level from 13 TeV LHC data with an integrated luminosity of $\mathcal{L}=140$ fb$^{-1}$, whereas HL-LHC (with 3 ab$^{-1}$ integrated luminosity) can probe up to 2.5 TeV. 
{The minimal constraints without assuming additional flavor structures from indirect high-$p^{}_{\rm T}$ searches of $q \bar{q} \to \mu^+ \mu^-$ are less competitive \cite{ Greljo:2017vvb, Angelescu:2021lln, CMS:2019tbu}.} All these constraints are summarized in Fig.~\ref{fig:contours_LQ}.

\vspace{0.05 in} 
\prlsection{Implications  of  $\mathbold{R_{K^{(*)}}}$ anomaly at a muon collider}{.}%
The transition of $b \to s \mu^+ \mu^-$ in meson decays is 
directly applicable in a muon collider via $\mu^+ \mu^- \to b \bar s$. This simple two-body scattering allows to directly test any explanation for the anomalous $R_{K^{(*)}}$ ratios, and we utilize it to study the sensitivity on the representative explanations of the anomalies, i.e.\ a $Z'$ and scalar as well as vector leptoquarks.

The Feynman diagrams of the relevant processes are depicted in Fig.~\ref{fig:feyn}. For the $Z'$ model, we have an $s$-channel process, and a resonance enhancement is available when the center-of-mass energy $\sqrt{s}$ is near the $Z'$ mass $M^{}_{Z'}$. In contrast, the $S^{}_3$ leptoquark  mediates a $t$-channel process. 

Besides the explicit realization of the cross section, we can describe the situation in an effective language. When the $Z'$ or leptoquark mass is larger than the 
center-of-mass energy, the operators with coefficients $C^{\mu}_9$ and $C^{\mu}_{10}$ are 
responsible for the transition. 
The cross section of $\mu^+ \mu^- \to b  \bar s$  is then 
\begin{eqnarray}
	\sigma(s) =  \frac{G^{2}_{\rm F} \alpha^2 |V^{}_{tb} V^{*}_{ts}|^2s}{8 \pi^3}\left( |C^{\mu }_{9}|^2 + |C^{\mu }_{10}|^2\right) .
\end{eqnarray}
Taking the best-fit scenario of $B$ anomalies, $ C^{\mu}_{9} = - C^{\mu}_{10} = -0.43$, we obtain the event number of $bs$ final states $\sigma(s)\cdot L$ ($L$ being the luminosity) as
\begin{eqnarray}
\#{\rm signal} \simeq 10^{3} \left(\frac{\sqrt{s}}{6~{\rm TeV}}\right)^2 \left(\frac{L}{4~{\rm ab^{-1}}} \right) . \notag
\end{eqnarray}
As a naive comparison, we obtain the relevant SM background in the form of quark dijets (ignoring flavor tagging, see below), which turns out to be 
$1.2\times 10^{5} \cdot (6~{\rm TeV}/ \sqrt{s})^2 \cdot (L/4~{\rm ab^{-1}})$. The signal exceeds the fluctuation of SM background at around $3\sigma$ level, which  is very encouraging. The signal-to-background ratio is roughly proportional to ${s}^2$; therefore to enhance the sensitivity to the effective operators,  larger $\sqrt{s}$ is preferred. 
{ With $\sqrt{s} = 10~{\rm TeV}$ and $L= 10~{\rm ab^{-1}}$,  values  of $|C^{\mu}_{9}| = |C^{\mu}_{10}|$ as small as $0.16$ can be reached at $3\sigma$ level, which covers the $2\sigma$ range of $|C^{\mu}_{9}| = |C^{\mu}_{10}| \in [0.29,0.57]$ even without the flavor tagging.  
For comparison, the current LHC (projected HL-LHC) limit on the coefficients of effective operators reads $|C^{\mu}_{9}| = |C^{\mu}_{10}| < 100~(39)$~\cite{Greljo:2017vvb}. These hadron collider bound on the effective operators is set by searching for the high-$p^{}_{\rm T}$ tails of the dimuon spectrum, which is not as efficient as a muon collider. In the supplemental material we discuss more details on the muon-collider sensitivity on effective operators.} 
Before discussing the explicit realizations of the process, we consider  general background issues.

The dijet signal of the $bs$ final state is  contaminated by $\mu^+ \mu^- \to j  j$, where $j$ can be $u$, $d$, $s$, $c$ and $b$, due to imperfect flavor reconstruction. The sensitivity depends on the $b$-jet tagging efficiency as well as the mistag rate (identifying a light quark jet as a $b$-jet). In this work, we assume an  experimental configuration with a $b$-jet tagging efficiency $\epsilon^{}_{b} = 70\%$~\cite{Liu:2021jyc} and  mistag rates $\epsilon^{}_{uds} = 1\%$ for light quarks and $\epsilon^{}_{c} = 10\%$ for $c$ quarks~\cite{Aaboud:2018xwy,Auerbach:2014xua,Jamin:2019mqx}.
We require in our analysis that one jet is tagged as a $b$ jet, while the other is not. 
We continue with some comments on the backgrounds: 
\begin{itemize}[noitemsep,topsep=1pt,leftmargin=5.5mm]
	\item $\mu^+ \mu^- \to u \bar{u},d \bar{d},s \bar{s},c \bar{c}$:  With the tagging requirement, the total cross section for these processes will be reduced by a factor of $2\epsilon^{}_{uds,c} \cdot (1-\epsilon^{}_{uds,c})$, where the factor $2$ originates from  two choices of tagging. 
	\item $\mu^+ \mu^- \to b \bar{b}$:  To pass our event criteria, one $b$-jet is required not to be $b$-tagged, and the cross section is reduced by a factor $2\epsilon^{}_{b}\cdot (1-\epsilon^{}_{b})$. Note that  one could likely  further optimize the selection criteria until a higher signal-to-noise ratio is obtained. 
\end{itemize}
{
In addition, there could be background contributions  from top quarks. However, their identification  relies crucially on the tagging of a $b$ quark in their decay $t \to W b$. Above TeV energies, the top-antitop final states are highly boosted, such that multiple final jets may overlap~\cite{Lillie:2007yh}.
However, the fractional momentum carried by the $b$-tagged jet always lies below $\sqrt{s}/2$, which should be well separated from the prompt $b$ jet~\cite{Auerbach:2014xua} with proper energy cuts. Thus, in our analysis we assume the top background to be negligible. An inclusion should not affect our results much, because other dijet backgrounds remain dominant.}

For illustration, we will investigate three collider setups with center-of-mass energies and luminosities, namely $(\sqrt{s}, L)  = (3~{\rm TeV}, 1~{\rm ab^{-1}})$, $(6~{\rm TeV}, 4~{\rm ab^{-1}})$ and $(10~{\rm TeV}, 10~{\rm ab^{-1}})$. For completeness, results with and without  flavor tagging will be given. 
We will use FeynCalc~\cite{Mertig:1990an,Shtabovenko:2016sxi,Shtabovenko:2020gxv} and FeynArts~\cite{Hahn:2000kx} for the numerical calculations of the scattering amplitudes.

For simplicity, we perform the analysis at the parton level. An angular cut $ 10^{\circ}<\theta <170^{\circ}$ on the final state jets will be implemented. 
Since there are no divergent $t$-channel contributions, a slightly stricter or looser angular cut will not affect the final sensitivity much. The signal jets are monoenergetic with $E^{}_{j} = \sqrt{s}/2$. At the parton level, no additional cut on the energy needs to be considered.
The statistical significance is measured by 
\begin{eqnarray} \label{eq:chi2}
	\chi^2 = \sum^{}_{i}\frac{(N^{}_{i}-\widetilde{N}^{}_{i})^2}{N^{}_{i} + \epsilon^2 \cdot N^{2}_{i}}
	\;,
\end{eqnarray}
where $N^{}_{i}$ is the expected total event number of signal and backgrounds, and $\widetilde{N}^{}_{i}$ is the assumed event number observed by the experiment. 
The sensitivity can be generated by setting $\widetilde{N}^{}_{i}$ to be SM backgrounds only, i.e.\ a null signal. Further,  
$\epsilon$ denotes the possible systematic uncertainty, which will be fixed as  $0.1\%$~\cite{Han:2020uak} in our work. 
Setting $\epsilon$ to a higher value of $1\%$, which is comparable to the signal-to-background ratio without flavor tagging (see the supplemental material for more details), will dilute the significance. Nevertheless, after the flavor tagging procedure, the effect of systematic uncertainty is not significant as long as $\epsilon$ stays below  $2\%$.
The index $i$ sums over polar angles, for which we take a bin-size of  $\cos{\theta}$ as $0.1$. We highlight that we have checked that a finer binning does not improve the significance much, as the spectrum shape is already well contained with our choice.

The final sensitivity contours are shown in Fig.~\ref{fig:contours_Zp} (for $Z'$) and Fig.~\ref{fig:contours_LQ} (for leptoquark), where the red, blue and green curves correspond to the muon collider setups $(\sqrt{s}, L)  = (3~{\rm TeV}, 1~{\rm ab^{-1}})$, $(6~{\rm TeV}, 4~{\rm ab^{-1}})$ and $(10~{\rm TeV}, 10~{\rm ab^{-1}})$, respectively. The solid (or dashed) curves stand for the case without (or with) flavor tagging.
For comparison, the parameter regions explaining the $R^{}_{K^{(*)}}$ anomaly for $Z'$ and leptoquark models are given as yellow bands. 
For the $Z'$ case, there is a resonance near the center-of-mass energy. 
The width of the resonance depends on the couplings $\lambda^{\rm L}_{22}$ and $\lambda^{\rm Q}_{23}$ via $\Gamma = (2|\lambda^{\rm L }_{22}|^2 + 3|\lambda^{\rm Q }_{23}|^2) /(12\pi)$.
For small $\lambda^{\rm Q }_{23}$, the width is dominated by our choice of $\lambda^{\rm L }_{22}$.
If the ${Z'}$ and leptoquark masses are much smaller than the center-of-mass-energy, the sensitivity curves do not depend on the mediator mass. In this case, since the collider setups have luminosities $L \propto \sqrt{s}^2$,  the event number $\sigma(s) \cdot L$ will be a constant for $\sigma(s) \propto s^{-1}$ at large momentum transfer. 
At large $Z'$ and leptoquark masses, the mediator is decoupled, and the contours of the two models converge to each other. We note that in this regime the results will  be applicable to any effective theory described by  Eq.~{(\ref{eq:Leff})}. 
Some further comments are in order:
{
\begin{itemize}[noitemsep,topsep=1pt,leftmargin=5.5mm]
    \item Due to the constraints from neutrino trident production and $B^{}_{s}$ mixing, the parameter space is very limited for the $Z'$ scenario.
    The coverage of parameter space by the muon collider depends on the value of $\lambda^{\rm L}_{22}$. 
    It is worth noting that the dimuon signal from $\mu^+ \mu^- \to \mu^+ \mu^-$ is able to cover all the $\lambda^{\rm L}_{22}$ and $M_{Z'}$ values explaining the $B$ anomalies~\cite{Huang:2021nkl}. In this case,  the inclusion of $\mu^+ \mu^- \to b \bar s$ helps to clarify that the new physics is indeed what causes the $B$ anomalies.
    The $B^{}_{s}$ mixing data prefers larger $\lambda^{\rm L}_{22}$ values. If we take $\lambda^{\rm L}_{22} = 1$, a window between the projection for the  HL-LHC and the muon collider setup with $\sqrt{s} = 3~{\rm TeV}$ may survive. But this window is expected to be covered by means of radiative return, i.e., $\mu^+ \mu^- \to b s \gamma$. For the extreme case $\lambda^{\rm L}_{22} = \sqrt{4\pi}$ where more parameter space is valid to explain the $R_{K^{(*)}}$ anomaly, the muon collider with $\sqrt{s} = 6~{\rm TeV}$ will rule out most of the favored parameter space. Combining the HL-LHC and the muon collider sensitivities we observe that there is still a corner of the parameter space left.
    \item For the case of leptoquarks, most of the parameter space will be probed with the muon collider $\sqrt{s} = 6~{\rm TeV}$ and $L = 4~{\rm ab^{-1}}$.
    Only a tiny window around  $3~{\rm TeV}$ for scalar-leptoquarks may survive, which can be of course covered by a larger integrated luminosity (e.g., with $L = 16~{\rm ab^{-1}}$ for $\sqrt{s} = 6~{\rm TeV}$ no space will be left). {The parameter space of vector leptoquark and coefficients of effective operators can be fully covered with $\sqrt{s} = 6~{\rm TeV}$ and $L = 4~{\rm ab^{-1}}$.}
\end{itemize}
}

\vspace{0.05 in} 
\prlsection{Conclusion}{.}%
Processes with muons are a reliable source of anomalies which could lead to the discovery of long-awaited new physics beyond the Standard Model. A muon collider is then an ideal machine to probe these effects further. Here we have focused on the highly interesting $R_K$ and $R_{K^\ast}$ ratios, which are object to intense studies in terms of heavy $Z'$ bosons and leptoquarks. 
We have demonstrated  that the parameter space of such models can be mostly covered at currently discussed muon collider setups, which adds exciting physics potential to these facilities. \\


\appendix
\vspace{-0.2cm}
\begin{widetext}
\section{Supplemental Material}
\subsection{Signal-to-background Ratio}
In Fig.~\ref{fig:SNR}, an illustration of the signal-to-background ratio before (solid) and after (dashed)  flavor tagging is given for the leptoquark model with parameter sets explaining the $B$ anomalies $y^{\rm LQ}_{23} \cdot y^{\rm LQ}_{22} = 0.02$ and $M^{}_{S_3} = 5~{\rm TeV}$ (red curves), as well as  $y^{\rm LQ}_{23} \cdot y^{\rm LQ}_{22} = 0.7$ and $M^{}_{S_3} = 30~{\rm TeV}$ (blue curves). One can observe that with our  flavor tagging assumptions, the signal-to-background ratio can be enhanced by one order of magnitude. Considerable variations of the signal-to-background ratio over the polar angle can be noticed for the case with $M^{}_{S_3} = 5~{\rm TeV}$, which helps to preserve the statistical significance against a possibly large systematic uncertainty. 
\begin{figure}[b]
	\begin{center}
		\includegraphics[width=0.5\textwidth]{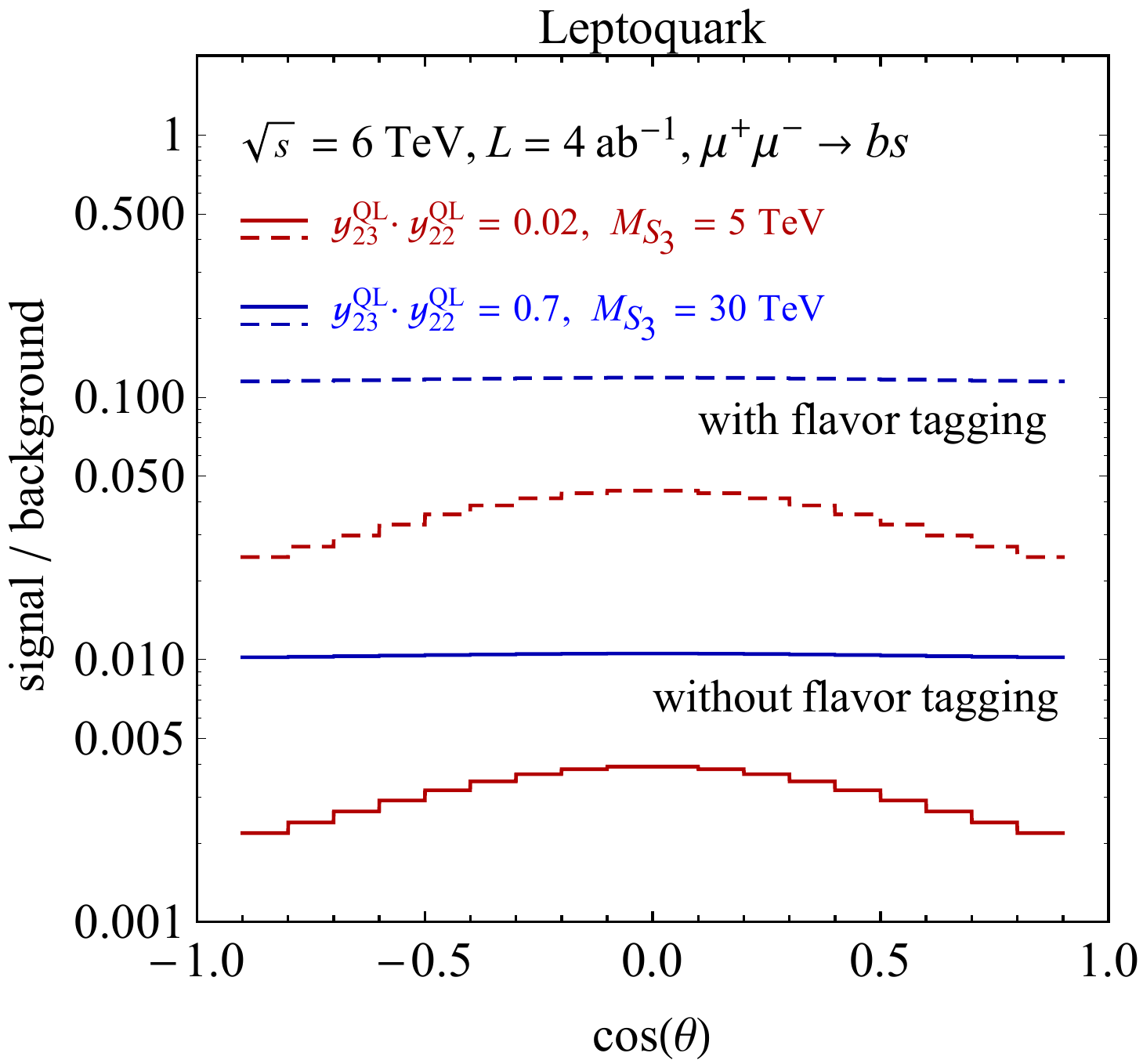} 
	\end{center}
	\caption{The ratio of leptoquark signal to SM background at a muon collider with  $\sqrt{s} = 6~{\rm TeV}$ and $L=4~{\rm ab^{-1}}$. The red curves stand for the scenario $y^{\rm LQ}_{23} \cdot y^{\rm LQ}_{22} = 0.02$ and $M^{}_{S_3} = 5~{\rm TeV}$, where the leptoquark mass is comparable to the collision energy. The blue curves stand for the scenario $y^{\rm LQ}_{23} \cdot y^{\rm LQ}_{22} = 0.7$ and $M^{}_{S_3} = 30~{\rm TeV}$. Here the leptoquark can safely be integrated out, and the scattering is described by  effective operators. The case with (without) flavor tagging is shown as dashed (solid) curves.
	}
	\label{fig:SNR}
\end{figure}
However, for the case with $M^{}_{S_3} = 30~{\rm TeV}$, the leptoquark is basically decoupled, and the signal-to-background ratio is nearly a constant if we do not distinguish quark and antiquark. A possible tagging of the $b$ quark charge~\cite{TheATLAScollaboration:2015ggd} will distort the flat signal-to-background ratio.

\subsection{Vector Leptoquark}
The Lagrangian describing the $U^{}_{1}$ vector-leptoquark reads
\begin{equation}
	\mathcal{L}_{U_{1}}=-M_{U_{1}}^{2}\left|U_{1}\right|^{2}+y_{i \alpha}^{\rm LQ} \overline{Q^{i}} \gamma^{}_{\mu} L^{\alpha} U_{1}^{\mu}+\mathrm{h.c.}
\end{equation}
The corresponding contribution to the Wilson-coefficients at tree-level is similar to the $S^{}_{3}$ leptoquark, namely
\begin{equation} 
	C_{9}^{\mu}=-C_{10}^{\mu}=\frac{\pi}{\sqrt{2} G_{\rm F} M_{U_{1}}^{2} \alpha}\left(\frac{y_{32}^{\rm LQ} y_{22}^{\rm LQ *}}{V_{t b} V_{t s}^{*}}\right). 
\end{equation}
When the leptoquark mass is much larger than the colliding energy, the effects induced by $S^{}_{3}$ and $U^{}_{1}$ leptoquarks at muon colliders will be indistinguishable. 
When the leptoquark mass is negligible compared to the colliding energy, the $t$-channel exchange of $U^{}_{1}$ leptoquark will enhance the cross section significantly by a factor of $1/(Q^2+M_{U_{1}}^{2})^2$.
However, for the $S^{}_{3}$ case, the scalar coupling, which reverses the chirality, does not feature a $t$-channel enhancement. This can be easily seen: the vertex for the scalar coupling contributes a factor ${\rm Tr}(\slashed{p} \slashed{k}) = 4 p \cdot k  \propto Q^2$ with $p$ and $k$ being the four momentum of initial and final fermions coupled to the leptoquark, and the $t$-channel enhancement when $Q^2 \to 0$ is therefore canceled. 
As a consequence, in Fig.~3 of the main manuscript, we have better sensitivities at small masses for the vector leptoquark.

\subsection{Sensitivity to Effective Operators}
In Fig.~\ref{fig:effoper}, we show the $3\sigma$ sensitivity of muon colliders to $|C^{\mu }_{9}|^2 + |C^{\mu }_{10}|^2$ as a function of the colliding energy $\sqrt{s}$. The yellow band corresponds to the $2\sigma$ range favored by the global analysis, namely $C^{\mu}_{9} = -C^{\mu}_{10} \in [0.29,0.57]$.
The blue region (dashed blue curve) shows the excluded values of $|C^{\mu }_{9}|^2 + |C^{\mu }_{10}|^2$ for a given colliding energy $\sqrt{s}$ assuming only the Standard Model background is observed without (with) flavor tagging.
With the setup $\sqrt{s} = 6~{\rm TeV}$ and $L= 4~{\rm ab^{-1}}$, the best-fit point $ C^{\mu}_{9} = - C^{\mu}_{10} = -0.43$ can be reached without the flavor tagging.
We note that with the colliding energy $\sqrt{s} \gtrsim 6~{\rm TeV}$ and the flavor tagging the entire $2\sigma$ range of parameter space favored by the $B$ anomalies can be covered.
\begin{figure}[h]
	\begin{center}
		\includegraphics[width=0.5\textwidth]{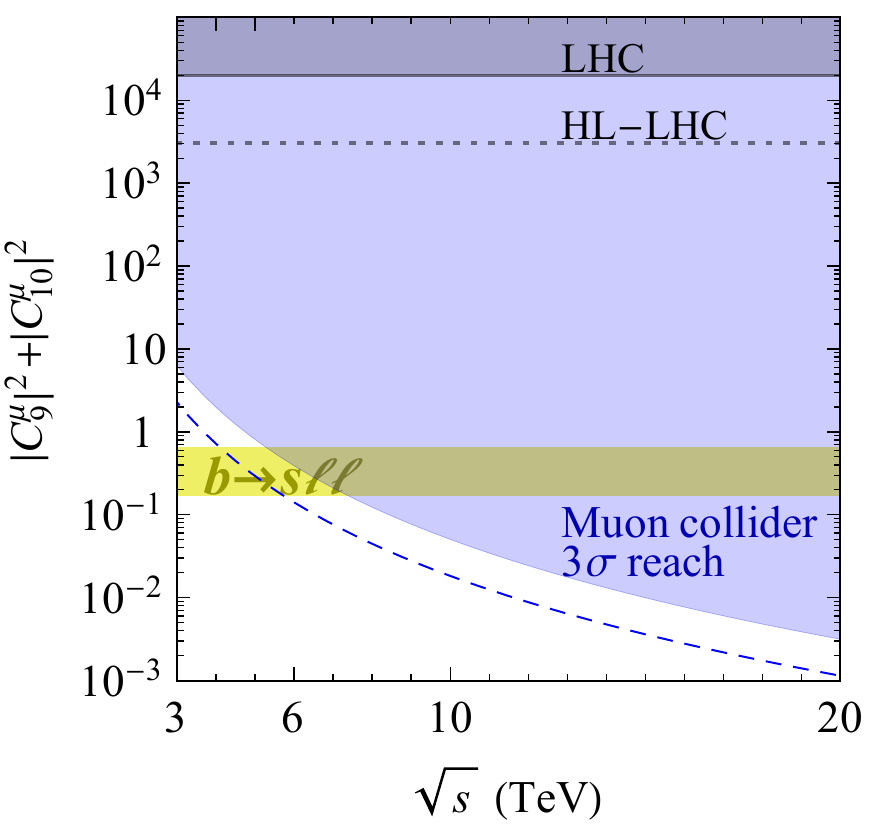} 
	\end{center}
	\caption{The sensitivity of muon colliders to the square sum of effective operator coefficients $|C^{\mu }_{9}|^2 + |C^{\mu }_{10}|^2$ as a function of the colliding energy $\sqrt{s}$. The luminosity has been assumed to satisfy the benchmark value $L = 4~{\rm ab^{-1}} \cdot [\sqrt{s}/(6~{\rm TeV})]^2$. The blue region (the dashed blue curve) is the $3\sigma$ exclusion parameter space assuming no excess beyond the Standard Model background has been observed without (with) the flavor tagging, while the yellow band indicates the $2\sigma$ range favored by the global fit of $B$ anomalies assuming $C^{\mu}_9 = - C^{\mu}_{10}$~\cite{Altmannshofer:2021qrr,Aebischer:2019mlg}. The LHC limit and the HL-LHC projection by looking for high-energy dimuon tails, assuming only the $bs \mu\mu$ couplings, are given as black shaded region and dotted line, respectively~\cite{Greljo:2017vvb}.
	}
	\label{fig:effoper}
\end{figure}
\end{widetext}


\prlsection{Acknowledgments}{.}%
GYH was supported by the Alexander von Humboldt Foundation. FSQ is supported by the Sao Paulo Research Foundation (FAPESP) through grant 2015/158971, ICTP-SAIFR FAPESP grant 2016/01343-7, CNPq grants 303817/2018-6 and 421952/2018-0, and the Serrapilheira Institute (grant number Serra-1912-31613).

\bibliographystyle{utphys}
\bibliography{reference}

\end{document}